\documentclass{aa}

\makeatletter
\renewcommand*\aa@pageof{, page \thepage{} of \pageref*{LastPage}}
\makeatother

\usepackage{hyperref}
\usepackage[dvipsnames]{xcolor}
\hypersetup{colorlinks=true, 
            citecolor=Cerulean, 
            linkcolor=Cerulean, 
            urlcolor=Cerulean}
\usepackage{graphicx}
\usepackage{wrapfig}
\usepackage[rightcaption]{sidecap}
\usepackage{pdflscape}
\usepackage{txfonts}
\usepackage{siunitx}
\usepackage{amsmath}
\usepackage{amssymb}
\usepackage[normalem]{ulem}
\usepackage{lipsum}  
\usepackage{breqn} 

\usepackage[title]{appendix}

\usepackage{natbib}
\usepackage[switch]{lineno}

\begin{document}

\title{An open-access web tool for light curve simulation and analysis of small Solar System objects}

\author{
J. L. Rizos\inst{1}
\and
J. L. Ortiz\inst{1}
\and
P. J. Gutierrez\inst{1}
\and
I. M. Navajas\inst{1}
\and
L. M. Lara\inst{1}
}

\institute{
Instituto de Astrof\'isica de Andaluc\'ia – Consejo Superior de Investigaciones Cient\'ificas (IAA-CSIC), Glorieta de la Astronom\'ia S/N, E-18008, Granada, Spain.~~\email{jlrizos@iaa.es}
}

\date{Accepted for publication: October 3, 2025}

\abstract
   {Rotational light curves of small Solar System bodies provide key insights into their shapes, spin states, and surface properties. The generation of synthetic light curves based on shape models and photometric functions can be a powerful tool to interpret observational data and test hypotheses about the physical characteristics of these bodies.}
   {This work presents a web-based application designed to simulate rotational light curves of small airless Solar System bodies under user-defined geometrical and physical configurations, and it includes a dedicated module that generates the silhouette of a body at the epoch of a predicted stellar occultation, enabling direct comparison with observed chords.}
   {The application, developed in Python and Django, incorporates physical and empirical photometric models. It allows users to define viewing and illumination geometry, surface properties, and shape models (mesh files). A validation section enables comparison between simulation and observational data.}
   {The tool was validated using well-known objects such as (136108) Haumea, (101955) Bennu, and (433) Eros, for which projected silhouettes and synthetic light curves were generated using published shape and spin data. The results show excellent agreement with observations, confirming the reliability of the simulation engine. The application also allows users to explore surface heterogeneity, tumbling scenarios, or phase-angle dependencies.}
   {This platform offers a flexible and accessible framework for simulating and interpreting light curves across a wide variety of small airless Solar System bodies. Its modular design and planned extensions make it a promising tool for both current observational campaigns and future mission support.}

\keywords{
light curves -- photometric models --  web application -- software tools -- small Solar System bodies -- surface properties
}

   \maketitle

\section{Introduction}

Inferring the shape of a Solar System body from its rotational light curve is a well-established technique (e.g. \cite{Ostro1984,Ostro1988,Kaasalainen2001a,Kaasalainen2001b,Durech2010} -  DAMIT\footnote{\url{https://damit.cuni.cz/projects/damit/}}) typically relying on observations obtained over a range of viewing geometries and with sufficient signal-to-noise ratio. However, most inversion tools developed for this purpose are based on the assumption that the target body is convex, meaning that its surface lacks significant concavities or self-shadowing features. This assumption is clearly violated in the case of irregular or bilobate bodies such as 67P/Churyumov-Gerasimenko, (216) Kleopatra, or Arrokoth \citep{Preusker2017,Shepard2018,Keane2022}. Moreover, accurate modeling often depends on the availability of high-quality data collected from multiple observing geometries to constrain the three-dimensional shape effectively. However, these conditions are rarely met for distant bodies such as Centaurs or trans-Neptunian objects (TNOs); because of their great distance, the aspect angle remains nearly constant over time. In addition, their faintness often leads to low signal-to-noise ratios in the light curves. As a result, traditional shape-inversion techniques become inapplicable or unreliable in this context.

In this work, we adopt a different approach. Instead of inferring the shape from the light curve, we start from a predefined 3D shape model and compute its projection onto the sky plane and the rotational light curve, as would be observed from Earth, using its known orbital parameters and spin axis orientation (when available).

This tool was primarily designed to support the analysis of data obtained from stellar occultations, where the projected silhouette of a Solar System body on the sky plane can be directly constrained. To this end, the application incorporates a dedicated module that computes both the rotational phase and the projected silhouette at the epoch of the occultation, enabling direct comparison with observed chords. This functionality proves especially useful for visualizing and analyzing the set of rotational solutions that are consistent with the occultation profile. A recent application of this approach can be found in \cite{Rizos2024}. Nevertheless, the tool can also be employed in other planetary science contexts, such as studying the rotational properties and photometric behavior of minor bodies through light-curve observations.

While similar efforts have been presented in previous works--see DAMIT \citep{Durech2010} or \cite{Showalter2021}--our implementation introduces several novel features. First, it supports arbitrary non-convex 3D shapes with a high degree of topographic complexity, including user-supplied shape models. Second, our tool allows for the use of multiple photometric scattering models to compute the reflected light. Users can choose between empirical approaches, such as the Lommel-Seeliger or Minnaert laws, and more physically grounded models such as the Hapke formulation. Additionally, our application includes advanced options such as albedo variegation, axial precession, and self-shadowing, which further increase the physical realism of the simulations. The user can also upload their own observational data for direct comparison with simulated outputs. Another significant advantage of our tool is its accessibility: it is entirely web-based and requires no local installation. The user only needs to enter the desired parameters, with the option to upload their own 3D shape model and/or a rotational light curve for comparison. The server then performs the calculations, which are displayed in the interface and can be downloaded. Hosted on the servers of the Instituto de Astrofísica de Andalucía (IAA-CSIC), the application can be run remotely through an intuitive browser interface\footnote{Accessible at \url{https://lctsb.scitechss.iaa.es}}. To guide new users, a brief online manual is available under the title \emph{How does this work?} in the web interface, which provides practical instructions for basic usage without requiring a full reading of this article.

The structure of this paper is as follows. In Section \ref{sec:architecture}, we describe the overall system architecture, including the main components of the web application. Section \ref{sec:features} outlines the functionalities offered by the tool and explains how users can interact with the different simulation options. In Section \ref{sec:validation}, we present a validation cases, comparing simulations with real observational data. Finally, Section \ref{sec:conclusions} summarizes the conclusions.

\section{System architecture and implementation}\label{sec:architecture}

The tool is implemented as a web application built using the Django framework\footnote{\url{https://www.djangoproject.com/}}, a high-level Python-based web platform designed for rapid development and clean design. The architecture follows the Model–View–Template (MVT) pattern, which allows for a clear separation between the application logic, data structure, and presentation layer.

The interface is based on a series of HTML templates, which are dynamically rendered through Django views. These views serve as controllers that handle user input, validate parameters, and initiate the corresponding simulation processes on the server side. Once a simulation is executed, the results are returned to the frontend and displayed to the user. 

The backend computation is entirely written in Python. The user provides input parameters such as the shape model, spin axis orientation, or scattering law through the web interface, and all operations are handled server-side, ensuring consistency and reproducibility.

The application is deployed on a dedicated Linux server hosted at the Instituto de Astrofísica de Andalucía (IAA-CSIC). The system is optimized for consistent performance and direct access to hardware resources, without the use of virtualization or containerization layers.

\begin{figure}[ht]
    \centering
    \includegraphics[width=0.3\textwidth]{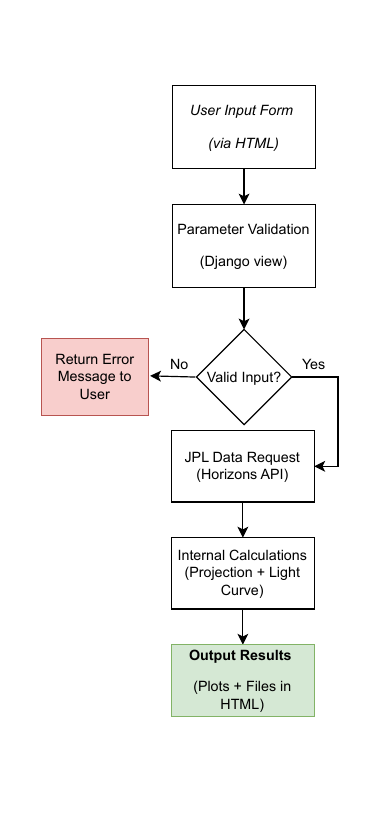}
    \caption{Workflow of the web application from user input to result rendering.}
    \label{fig:workflow}
\end{figure}

The interaction between the user and the web application follows a well-defined workflow, as illustrated in Figure \ref{fig:workflow}.The process begins when the user accesses the web interface and enters the required parameters. Upon submission, the data are passed to a Django view, which first performs a validation step (verifying that the object identifier, latitude and longitude are within acceptable ranges, etc.). If any inconsistency is detected, the view returns an error message to the user through the same web interface, highlighting the problematic field(s).

If the input passes validation, the view performs automated queries to the JPL Horizons system\footnote{\url{https://ssd.jpl.nasa.gov/horizons}} through its public batch interface. The tool retrieves the geometric parameters for the specified date without accounting for for the duration of the rotation period. For most small bodies this simplification is valid, but for very slow rotators (e.g., Apophis or Toutatis) users should simulate multiple dates to capture geometry changes. Once these data are retrieved, the server-side Python routines perform the relevant calculations: projection of the 3D model onto the sky plane, application of the selected photometric scattering law, and generation of the expected light curve.

The results—such as plots of the simulated light curve, sky-plane projection snapshots, and downloadable data files—are then passed back to the HTML template and rendered in the user’s browser. 

\section{Tool features and user configuration}\label{sec:features}

The web application provides two operational modes specifically designed for photometric studies of rotational light curves (the \textit{Known} and \textit{Generic body} modes) as well as an additional \textit{Occultation mode}, which projects the silhouette of a selected body at a given epoch, thereby enabling the analysis and planning of stellar occultations.

\subsection{Rotational light curves}
\begin{itemize}
\item Known body mode: This mode is designed for the simulation of Solar System objects with known rotation pole\footnote{The rotation pole coordinates must be provided in ecliptic coordinates, longitude from $0^\circ$ to $360^\circ$ and latitude from $-90^\circ$ to $+90^\circ$, as is the standard practice in this field. Once defined, the rotation sense follows the right-hand rule.} and orbital parameters. The user enters the Minor Planet Center (MPC) designation and the observation date. The tool generates both the rotational light curve and the projected view sky-plane as seen from Earth using the ephemeris data from the JPL Horizons system. 

\item Generic body mode: This mode enables simulations under user-defined geometric conditions. The observing and illumination geometries are specified directly by the user, including the phase angle, and the spin axis is defined within a custom reference frame. This is particularly useful for testing hypotheses or exploring different subsolar and subobserver configurations. It offers full flexibility for cases where the viewing geometry changes significantly over the course of a single rotation.
\end{itemize}

In both modes, users can provide a custom shape model in \texttt{.obj} format (with triangulated facets) or define a triaxial ellipsoid by specifying axes \emph{a}, \emph{b}, and \emph{c}. Shape models should be oriented such that the forward axis is \texttt{Y} and the up axis is \texttt{Z}. This ensures that the semiaxes \emph{a}, \emph{b}, and \emph{c} align with the \texttt{X}, \texttt{Y}, and \texttt{Z} axes, respectively. Users are encouraged to limit models (when possible) to fewer than 3000 facets for optimal performance. The internal ellipsoid is generated with 642 facets by default.

Several additional simulation features are available to users:
\begin{itemize}
\item Albedo spot: Enables the simulation of localized surface heterogeneities by allowing the user to define a specific surface region—based on the body's geocentric latitude and longitude—whose the Radiance Factor (RADF) is modified by a user-defined percentage.

\item Precession: Simulates the motion of the rotation pole as it precesses over time, instead of remaining fixed. The pole ($\lambda$ and $\beta_c$ represent longitude and co-latitude respectively) follows a conical trajectory, varying sinusoidally as a function of the rotational phase. This behavior is controlled by three user-defined parameters as follows:

\begin{subequations}
\begin{align}
\lambda(t) &= \lambda_0 + A_{\lambda}  \cos(2\pi \omega_p t) \\
\beta_c(t) &= \beta_{c0} + A_{\beta_c}  \sin(2\pi \omega_p t)
\end{align}
\end{subequations}

where $\lambda_0$ and $\beta_{c0}$ are the initial orientation of the rotation pole, $A_{\lambda}$ and $A_{\beta_c}$ are the amplitudes of the precessional motion, $\omega_p$ is the precession frequency (in cycles per rotation), and $t$ is the rotational phase (ranging from 0 to 1). This option enables the simulation of dynamic spin states and allows for the exploration of more realistic or perturbed rotational scenarios.

\item Self-shadowing: Enables the simulation of shadows cast by the object's own surface features, enhancing the physical realism of the reflected light distribution. For each facet, the algorithm checks whether the direction of incoming sunlight intersects any other facet of the shape model, thus determining whether the facet is illuminated or in shadow. This is implemented using the Möller–Trumbore ray-triangle intersection algorithm~\cite{moller1997}, which efficiently tests whether a ray defined by the Sun’s direction vector and the facet centroid intersects any of the mesh triangles. While this process increases computational time, it is essential for accurately modeling objects with significant concavities or complex topography. 

\end{itemize}

\subsubsection*{Photometric modeling}

In this tool, the RADF ---representing the ratio of the observed bidirectional reflectance to that of a perfectly diffuse surface illuminated at $i = 0^\circ$ \citep{Hapke2012}---can be computed in several ways.  We provide both empirical formulations and physically based models, allowing users to choose the level of complexity and realism appropriate for their application. In all cases, RADF depends on three photometric angles: the phase angle $\alpha$ (Sun-object-observer), the incidence angle $i$ (Sun-object surface normal), and the emission angle $e$ (observer-object surface normal).

Notice that this tool does not account for the object's distance to the Sun; instead, the solar constant is normalized to $S = 1$. Therefore, if absolute RADF values are needed, users must rescale the output accordingly. The value returned by the server corresponds to the integrated RADF as seen from Earth, computed by summing the contribution from each individual facet, as detailed below.

\paragraph{Empirical Models}

For empirical models, the RADF is approximated as the product of a phase function $f(\alpha)$ and a disk function $d(\alpha, e, i)$ as follows:

\begin{equation}
RADF = f(\alpha)\, d(\alpha, e, i)
\end{equation}

\subsubsection*{Phase Functions}

\begin{itemize}
\item Constant phase function: We provide this simplest case for users who do not need to account for any dependence on the phase angle, $f(\alpha) = 1$.

\item Minnaert phase function: This is a generalization of Lambert’s law, suggested by \cite{Minnaert1941} with the modification suggested by \cite{Takir2015}:

\begin{equation}
f(\alpha) = A_{\text{min}} \,\pi \,10^{-0.4(\beta \alpha + \gamma \alpha^2 + \delta \alpha^3)}
\end{equation}

where $A_{\text{min}}$, $\beta$, $\gamma$, and $\delta$ are user-defined parameters.

\end{itemize}

\subsubsection*{Disk Functions}

\begin{itemize}
\item Lambert: This model describes a perfectly diffuse (Lambertian) surface, which reflects incident light equally in all directions \citep{Lambert1760}. The reflected intensity is proportional to the cosine of the incidence angle, representing the projection of the incident flux onto the surface. The division by $\pi$ ensures energy conservation when integrating the radiance over the entire hemisphere. Although idealized, the Lambertian model is commonly used due to its simplicity and serves as a useful reference in photometric analyses:

\begin{equation}
d(i, e) = \frac{\cos i}{\pi}
\end{equation}
\\
\item Lommel-Seeliger: The Lommel-Seeliger law is based on a simple physical model of diffuse reflection under the assumption of single, isotropic scattering. It remains a valuable first-order approximation in planetary photometry and is still widely used in applications such as light curve inversion and shape modeling of minor bodies \citep{Fairbairn2005}. Unlike the Lambertian model, this formulation introduces an angular dependence on both incidence ($i$) and emission ($e$) angles, and it is particularly useful for modeling dark planetary surfaces or regolith-covered bodies such as asteroids or moons:

\begin{equation}
d(i, e) = \frac{\cos i}{\cos i + \cos e}
\end{equation}
\\
\item Minnaert: This disk function generalizes Lambert’s law by introducing the empirical exponent $k$, which accounts for non-uniform scattering characteristics and modulates the dependence on incidence ($i$) and emission ($e$) angles. The formulation adopted here is consistent with that introduced by \cite{Minnaert1941}:

\begin{equation}
d(i, e) = \cos^k i \cdot \cos^{k-1} e
\end{equation}

\end{itemize}

\paragraph{Hapke Model}

In addition to empirical formulations, we implement the Hapke model \citep{Hapke2012} which accounts for multiple scattering, shadow hiding, coherent backscatter, and macroscopic roughness. This model depends on nine physical parameters:

\begin{itemize}
\item $w$: single scattering albedo.
\\
\item $b_{\text{HG}}, c_{\text{HG}}$: Henyey-Greenstein parameters for the particle phase function.
\\
\item $B_{C0}, h_c$: coherent backscatter amplitude and width.
\\
\item $B_{S0}, h_s$: shadow hiding amplitude and width.
\\
\item $\theta_p$: macroscopic roughness slope angle.
\\
\item $\phi$: porosity factor.
\end{itemize}

The RADF is computed internally using the full formulation. For a complete detailed description of all the equations see Appendix \ref{appendix}.

\subsection{Occultation module}

This mode builds directly on the functionalities described above but is specifically designed for a direct application in the context of stellar occultations. Given the MPC designation of a target body, its sidereal rotation period (in hours), and the orientation of its spin axis, the tool computes the rotational phase for a user-specified date at which a stellar occultation is predicted, using the reference epoch $t_0$ (in JD) corresponding to a photometric rotational light curve maximum. As in the other modes, users may also upload a custom shape model in \texttt{.obj} format or define the principal axes of an ellipsoid with 642 facets by default.

A major advantage of this tool is that it enables the exploration of viewing-geometry variations across different dates. Since the on-screen output is rendered to scale, users can empirically determine the expected dimensions along any given direction and directly compare them with the observed occultation chords.

\subsection{Output}

The web interface provides a complete summary of all the parameters used in each simulation. If the \texttt{Generate images} checkbox is selected for Known and Generic modes, an animated GIF is displayed showing the projected shape of the object on the sky plane. For the Occultation mode, the interface also displays the rotational phase of the object at the inserted occultation date, and the corresponding projected figure. The colors of the surface facets reflect the solar incidence angle, with red indicating lower values (more irradiated areas) and blue indicating higher ones. The white arrow corresponds to the orientation of the rotation pole at the time of the observation. A scale bar is shown in the bottom-right corner. The default units of the tool are kilometers.

In the \textit{Known and Generic modes} we can see the synthetic light curve in relative magnitude which is computed as:

\begin{equation} \label{eq:rel_mag}
\text{Relative magnitude} = -2.5 \log_{10} \left( \text{RADF} / \overline{\text{RADF}} \right)
\end{equation}

where \text{RADF} is the sum of the radiance factor from all facets at a given rotational phase and $\overline{\text{RADF}}$ is the average radiance factor.

By clicking the \texttt{Download files} button, a ZIP archive is generated and downloaded to the user's local machine. This archive contains:
\begin{itemize}
  \item All generated images, each named with the corresponding rotational phase.
  \item The synthetic light curve as \texttt{synthetic\_LC.png}
  \item The shape model in \texttt{.obj} format.
  \item The animated projection GIF (when \texttt{Generate images} checkbox was selected).
  \item A text file named \texttt{output.dat} with the computed numerical results.
\end{itemize}
In the \texttt{output.dat} file, the first column lists the rotational phase, the second the RADF, and the third the relative magnitude as defined in Eq. \ref{eq:rel_mag}.

Similarly, in the case of the Occultation mode, the user can download a ZIP file containing the projected image for the occultation date, the shape model in \texttt{.obj} format, and the \texttt{output.dat} file with all the data corresponding to this case.

\section{Validation}\label{sec:validation}

\subsection{Projected shape}

First, we validate that the projected shape generated by the tool reproduces the actual silhouette of an object at a specific date using the \textit{Occultation mode}. For this test, we selected the trans-Neptunian dwarf planet (136108) Haumea, since projected occultation chords on the sky plane are available for January 21, 2017, when the body was observed near the absolute brightness minimum of its rotational phase \citep{Ortiz2017}. Haumea’s triaxial shape is well constrained ($a = 1161 \pm 30$ km, $b = 852 \pm 4$ km, $c = 513 \pm 16$ km), as well as its spin-axis orientation $(\alpha, \delta) = (285.1^\circ \pm 0.5^\circ,-10.6^\circ \pm 1.2^\circ$, J2000 equatorial), and rotation period ($3.915341 \pm 0.000005$ h) \citep{Ortiz2017}. From our light-curve database, we find a brightness maximum ($t_{0}$) at 2016-12-19 20:54:04.554 UTC, corresponding to JD 2457742.37088604.

\begin{figure}[ht]
    \centering
    \includegraphics[width=0.45\textwidth]{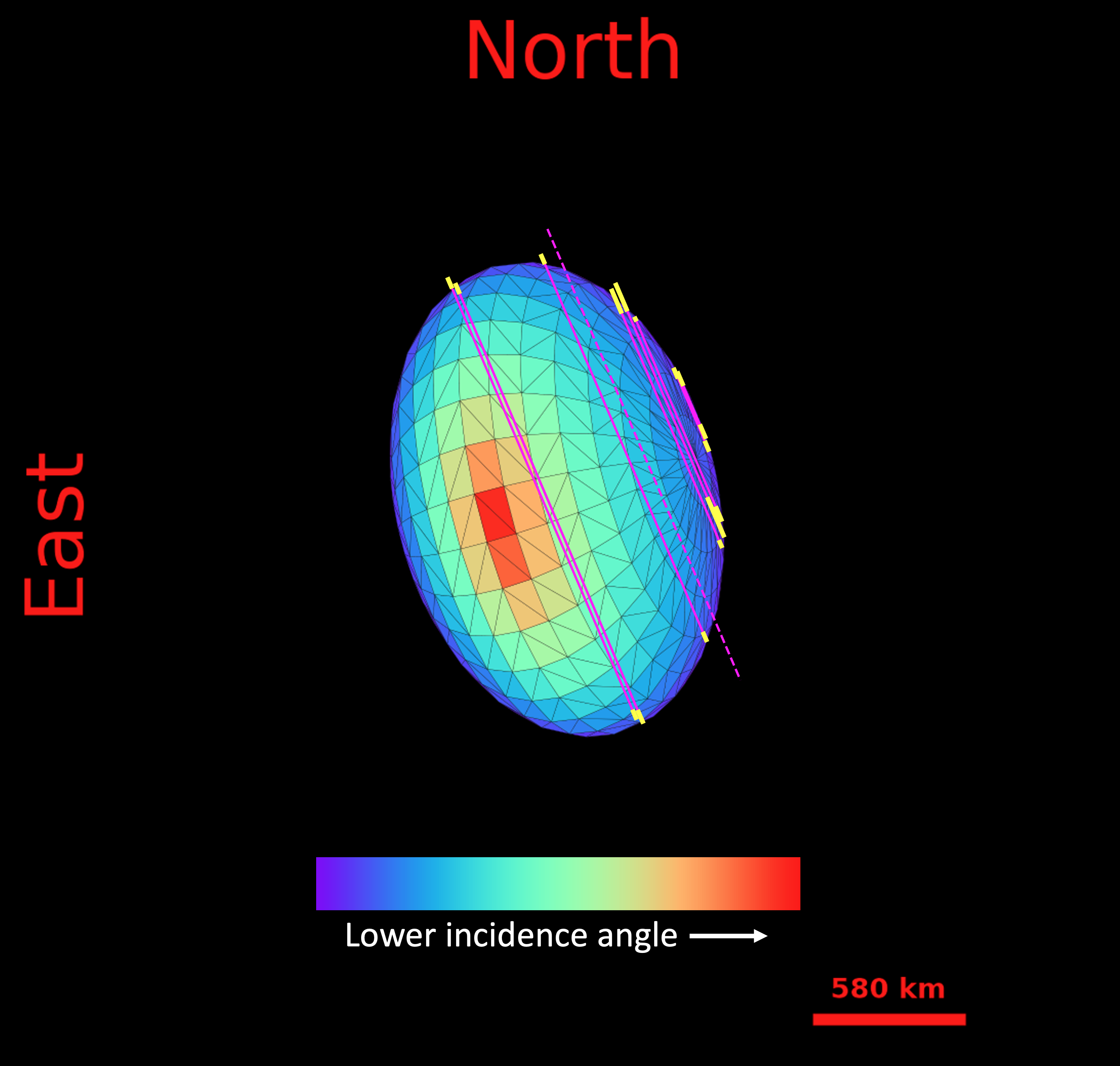}
    \caption{Haumea on January 21, 2017, as seen from Earth. On this date, \citet{Ortiz2017} reported observations from multiple Earth-based observatories of Haumea passing in front of a distant star (a multi-chord stellar occultation). These observations allowed them to constrain the shape of Haumea’s projected ellipse on the sky. The pink lines represent the projected occultation chords derived from the reported timings, while the yellow endpoints indicate the associated uncertainties in the ingress and egress times. The facet colors on the projected body indicate the relative level of irradiance received by each facet, with red corresponding to lower incidence angles and therefore higher irradiation. The overall agreement between the projected shape and the occultation chords supports the validity of our simulation.}
    \label{fig:haumea}
\end{figure}

The spin-axis orientation was then converted into ecliptic coordinates, as required by the tool (lon = $285.17^\circ$, lat = $12.05^\circ$), and we generated the projection for the 2017 occultation date. It yields Fig.~\ref{fig:haumea}, with a rotational phase of 0.75 from the LC maximum. We then extracted the stellar occultation timings from \citet{Ortiz2017}, projected the corresponding chords onto the sky plane, and overplotted them on the figure (pink and yellow lines; see also Fig.~2 in \citet{Ortiz2017}). This comparison shows the good agreement between the projected shape model and the observational data, supporting the consistency of our results.

\subsection{Rotational light curve}

To ensure that this validation is meaningful, it is necessary to have a highly accurate three-dimensional shape model of the object, a well-characterized photometric model, and observed light curves from Earth for comparison. For these reasons, we first selected asteroid (101955) Bennu as our validation target. This asteroid was orbited by the OSIRIS-REx spacecraft \citep{Lauretta2015}, which provided a high-resolution shape model. For this analysis, we adopt Bennu's shape model version 20, which features a spatial resolution of 75~cm and comprises approximately 200{,}000 triangular facets. The model is distributed in \texttt{.obj} format\footnote{\url{https://www.asteroidmission.org/updated-bennu-shape-model-3d-files/}}.

The mission also enabled a well-constrained characterization of Bennu's photometric properties for the four color filters of the OCAMS MapCam instrument: $b'$, $v$, $w$, and $x$, centered at 473, 550, 698, and 847 nm, respectively \citep{Rizk2018}. These filters were designed to match the passbands of the Eight-Color Asteroid Survey (ECAS) system \citep{Tedesco1982}. Conveniently, Bennu was also observed with the ECAS $w$ filter using the Kuiper 1.54 m reflector telescope in Arizona, USA \citep{Hergenrother2013}. A rotational light curve was constructed by  \citep{Hergenrother2013} from observations conducted between 14 and 17 September 2005. Accordingly, we simulated a synthetic rotational light curve for the same filter to ensure consistency.

The rotation pole of Bennu is located at $\alpha$ = $85.65^\circ$, $\delta$ = $-60.17^\circ$ \citep{Barnouin2019}, which corresponds to ecliptic coordinates of lon = $71.11^\circ$, lat = $-83.32^\circ$. Since Bennu is a near-Earth asteroid (NEA), its viewing geometry changes significantly over the course of the 4-day observational campaign. The solar phase angle for each observed day was $58.8^\circ$, $61.9^\circ$, $65.4^\circ$, and $69.4^\circ$, respectively. Therefore, it is necessary to take into account not only rotational variations but also phase angle changes. For this reason, we adopted the Minnaert photometric model, which incorporates both a phase function and a disk function, and whose parameters have been well constrained by \citet{Golish2021a}, as summarized in Table~\ref{tab:photometric_model_minnaert}.

\begin{table*}[ht]
\centering
\caption{Minnaert photometric model parameters for Bennu.}
\begin{tabular}{lccccc}
\hline
Filter &  $A_{\text{min}}$ & $\beta$ & $\gamma$ & $\delta$ & $k$   \\
\hline
w (698 nm) & 0.0132 & $3.610 \times 10^{-2}$ & $-2.966 \times 10^{-4}$ & $1.709 \times 10^{-6}$ & 0.6665  \\
\hline
\end{tabular}
\label{tab:photometric_model_minnaert}
\end{table*}

\begin{figure}[ht]
    \centering
    \includegraphics[width=0.4\textwidth]{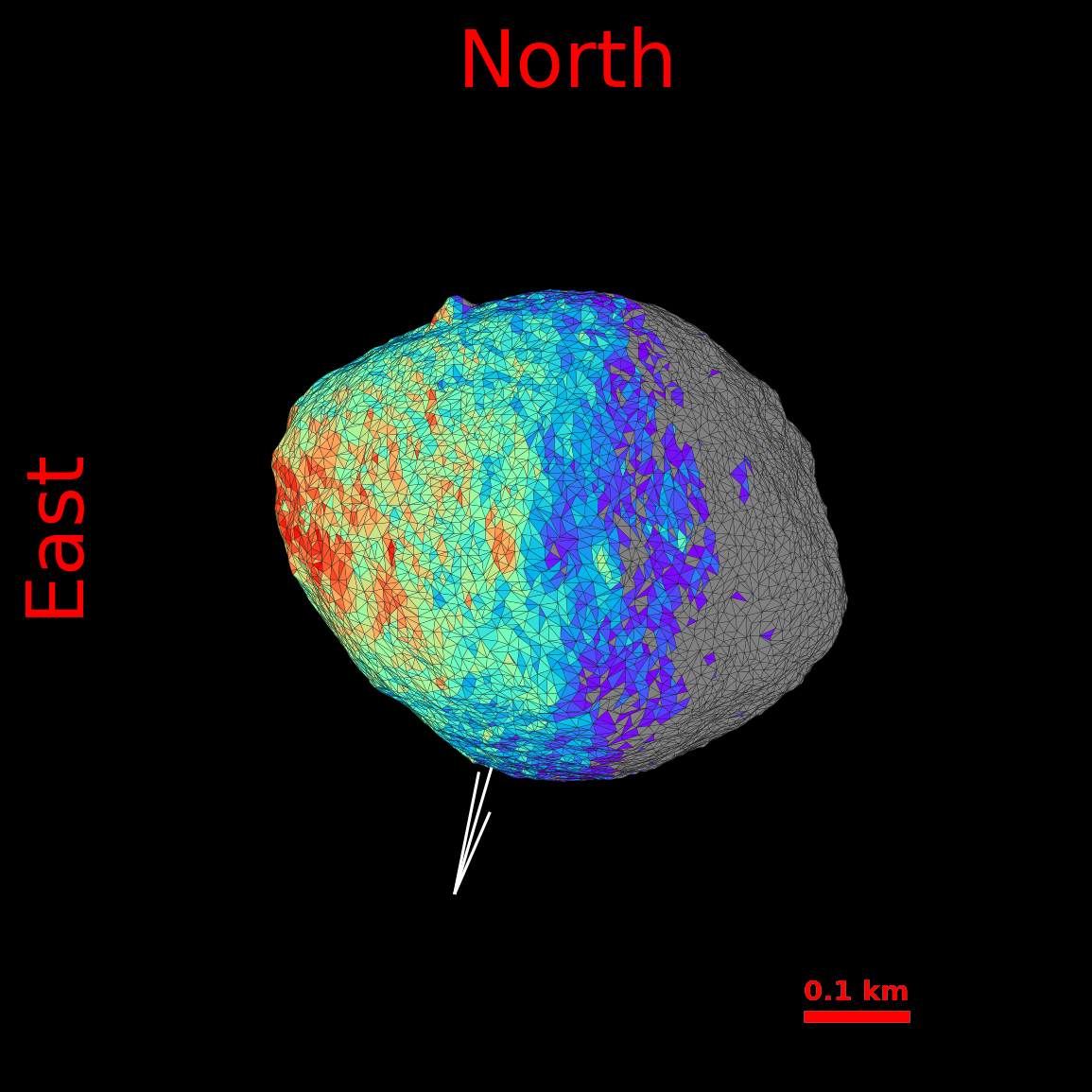}
    \caption{Projected view of Bennu, generated with a $\sim$20{,}000-facet downsampled shape model, as seen from Earth on 15 September 2005 (phase angle of $61.9^\circ$).}
    \label{fig:bennu}
\end{figure}

To reduce computation time, we downsampled the number of facets in the shape model by $10\%$. Figure \ref{fig:bennu} shows the projected shape of Bennu on 15 September 2025.

\begin{figure}[ht]
    \centering
    \includegraphics[width=0.5\textwidth]{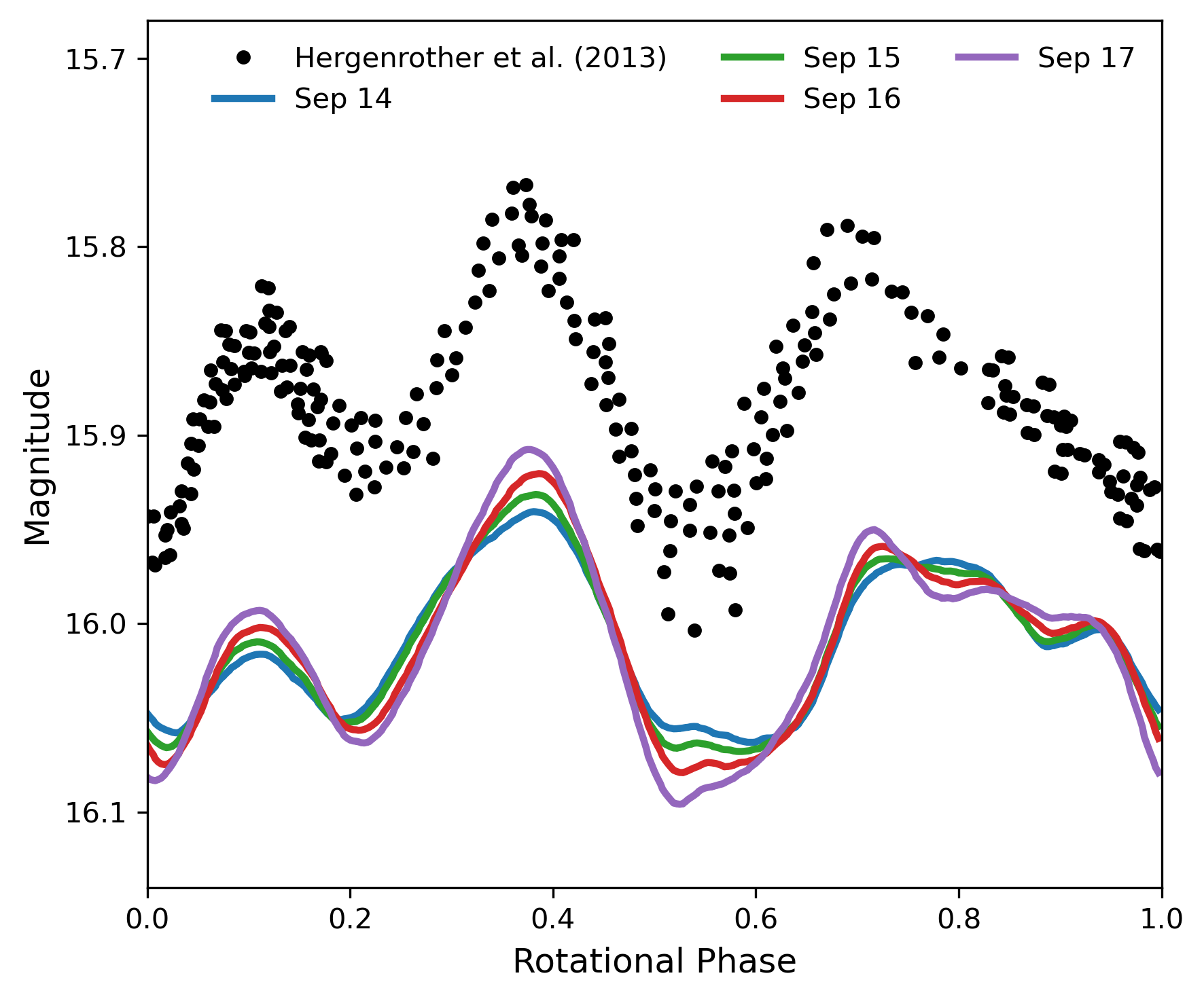}
    \caption{Black dots represent the observational data from the light curve published by \citet{Hergenrother2013}. Colored lines show the synthetic rotational light curves of Bennu as seen from Earth on 14–17 September 2005, sampled at one point per degree of rotation. An offset has been applied to synthetic light curves for clarity.}
    \label{fig:lc_comparison_bennu}
\end{figure}

In Figure \ref{fig:lc_comparison_bennu} we present the synthetic light curves generated for each day, with a sampling resolution of one data point per degree of rotational phase. To facilitate comparison for the reader, we extracted the observational data from the light curve published by \cite{Hergenrother2013} using the Automeris\footnote{\url{https://automeris.io/WebPlotDigitizer}} tool and plotted them together. An offset was applied to the synthetic curves for clarity.

The first result that highlights the usefulness of this tool is the identification of scatter in the observational light curve caused by the combination of multiple light curves affected by changes in viewing geometry---a dispersion that could otherwise be mistakenly attributed to photometric noise if unaccounted for.

It should be noted that, due to the complexity of this case, matching the observed photometric range required scaling the synthetic light curve amplitude by a factor of 1.5 — a correction that is justified for several reasons:

First, the photometric models provide a simplified description of surface scattering. It is well-known that photometric models tend to perform poorly for large values of phase, emission, or incidence angles \citep{Rizos2021}. In particular, regions near the limb, combined with limitations in the shape model, are expected to introduce biases. Furthermore, Bennu’s surface is covered with numerous boulders, some of which are of exogenous origin \citep{Tatsumi2021}, exhibiting albedo variations of up to $15\%$ \citep{Dellagiustina2019}. Each of these surface elements may contribute to distinct photometric behaviors that are not captured by the simulation. Finally, the shape model is inherently limited and does not fully capture the complexity of Bennu’s surface (see Fig \ref{fig:bennu_polycam}). To assess the sensitivity of the results to the shape model resolution, we repeated the same experiment by further downgrading the shape model by another $10\%$ (using instead only 2{,}000 facets). We observed a peak-to-peak amplitude variation of approximately $20\%$, underscoring the limitations imposed by model resolution.

\begin{figure}[ht]
    \centering
    \includegraphics[width=0.4\textwidth]{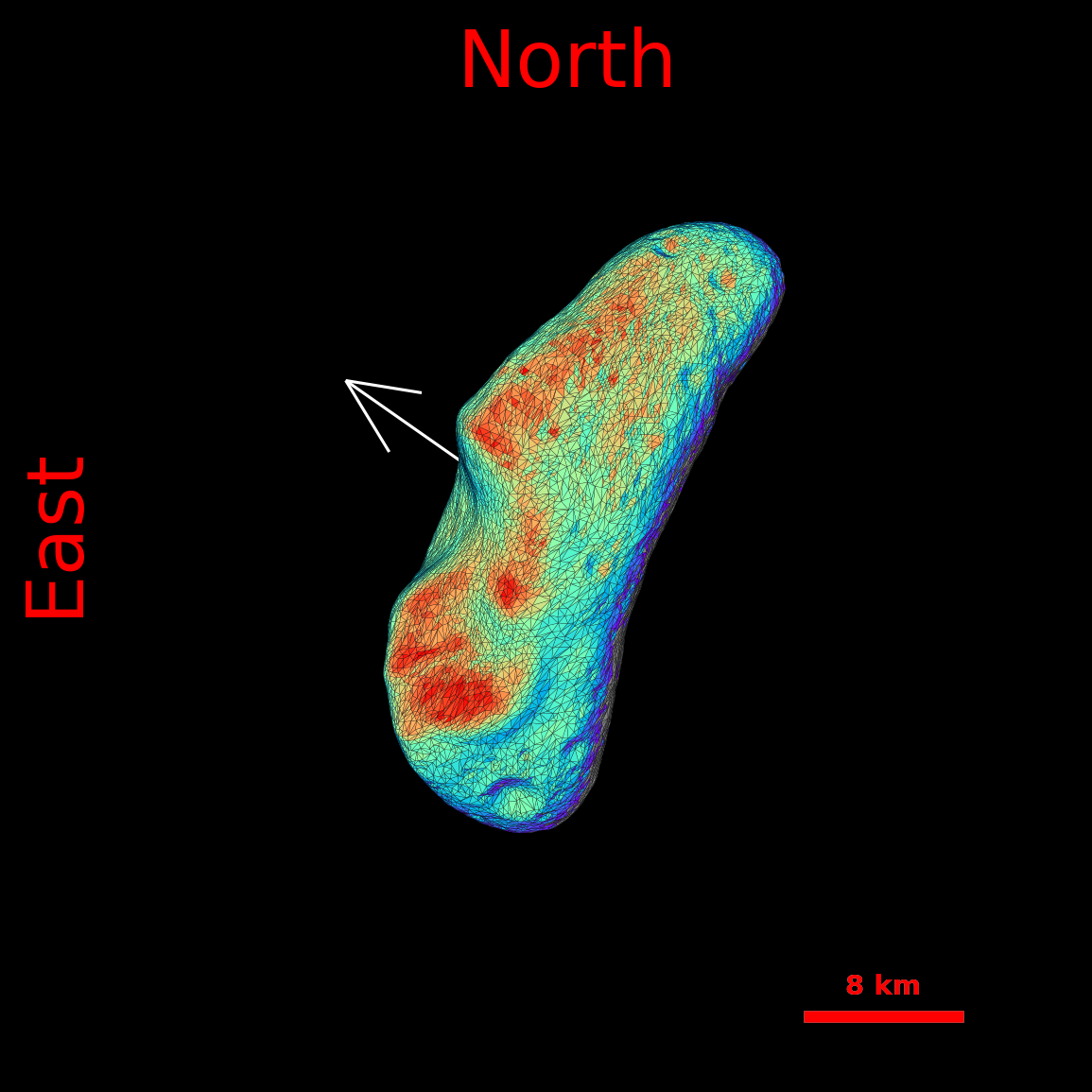}
    \caption{Projected view of Eros, generated with a  $\sim$20{,}000-facet downsampled shape model, as seen from Earth on 17 July 2016 (phase angle of $24.07^\circ$).}
    \label{fig:Eros}
\end{figure}

\begin{figure}[ht]
    \centering
    \includegraphics[width=0.5\textwidth]{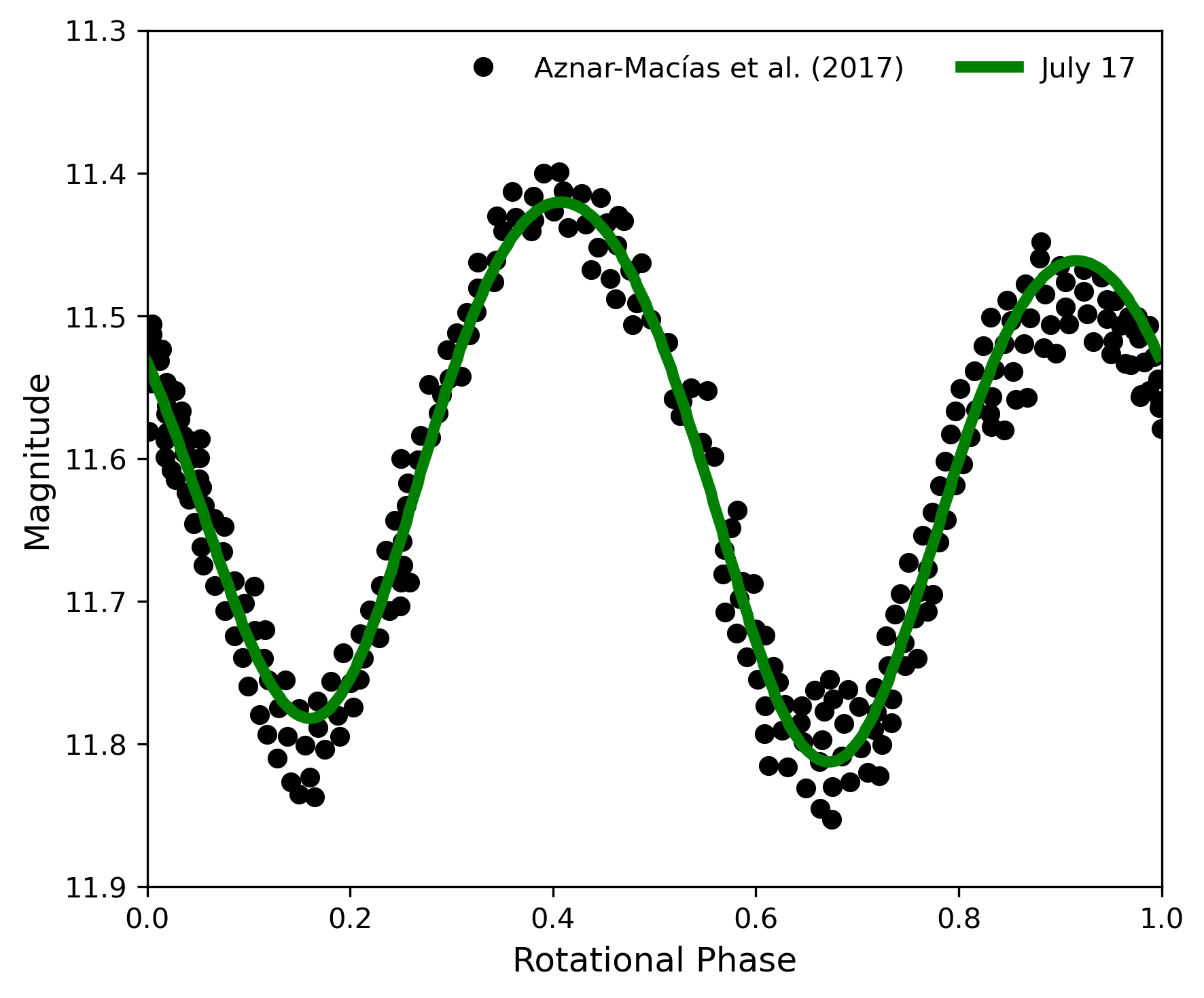}
    \caption{Black dots represent the observational data from the light curve published by \citet{Aznar2018}. Green line show the synthetic rotational light curve of Eros as seen from Earth on 17 2016, sampled at one point per degree of rotation.}
    \label{fig:lc_comparison_eros}
\end{figure}

We selected a second, less demanding example: asteroid (433) Eros. This object was observed by \citet{Aznar2018} on July 17, 18, 19, and 24, 2016, from the Aznar Observatory, located at the Centro Astronómico del Alto Turia in Aras de los Olmos, Spain. During this period, the solar phase angle varied from $24.0^\circ$ to $20.7^\circ$. Observations were conducted using a 0.36 m Schmidt-Cassegrain telescope equipped with a CCD camera and broad-band Astrodon Sloan r and J–V filters. The light curves were obtained over multiple nights and reduced to a common reference phase angle of $24.0^\circ$.
For comparison, we simulated a rotational light curve of Eros on July 17, 2016, when the observed phase angle matched $24.0^\circ$ (the spin axis of Eros in ecliptic coordinates is located at lon = $17.23 ^\circ$, lat = $11.34^\circ$ \citep{Yeomans2000}). Since no specific photometric model has been derived for Eros using these particular filters (it requires multi-epoch observations covering a wide range of phase and illumination angles with a consistent filter set, which is not the case here), we applied only disk function corrections using the Minnaert model, adopting a coefficient of $k = 0.7$, a value lying within the typical range reported for Solar System bodies (e.g., \cite{Li2005PhD,Jehl2008,Li2013,Masoumzadeh2015}). The shape model of Eros\footnote{\url{https://arcnav.psi.edu/urn:nasa:pds:gaskell.ast-eros.shape-model}} \citep{Gaskell2021Eros} was downsampled to  $\sim$20{,}000 facets to match the resolution used in the Bennu test case and reduce computation time.
Fig \ref{fig:Eros} shows the orientation of Eros as seen from Earth during the simulation, and Fig. \ref{fig:lc_comparison_eros} displays the synthetic light curve together with the observational data extracted using the Automeris tool. In this simpler case, the comparison did not require any scaling of the light curve amplitude.

\section{Conclusions}\label{sec:conclusions}

We have presented a web-based simulation tool for generating projected shapes and synthetic rotational light curves of small Solar System bodies. By leveraging high-resolution shape models, validated photometric laws, and realistic viewing geometries, the tool enables users to explore the physical properties and photometric behavior of asteroids, moons, and comets in a fully customizable framework.

As a validation case, we reproduced the silhouette of (136108) Haumea, whose shape is well constrained by multichord occultations, obtaining very good agreement. Then, we modeled the rotational light curves of (101955) Bennu and (433) Eros. Bennu is an almost spherical body (more precisely, diamond-shaped), with a light curve amplitude $\Delta m$ below 0.16 magnitudes. Therefore, this object constitutes a particularly stringent benchmark, pushing the tool to its sensitivity limits. The results demonstrate that the tool can reproduce subtle photometric variations under varying geometries, while also highlighting the limitations imposed by surface complexity and photometric modeling assumptions.

In summary, this tool provides a flexible and extensible framework for photometric modeling, with a focus on accessibility, scientific rigor, and future adaptability. It is our hope that it will contribute to the analysis and interpretation of small bodies observations across a wide range of planetary science applications.

\begin{acknowledgements}
We sincerely thank the anonymous referee for their constructive suggestions, which helped improve the clarity and quality of this manuscript. J. L. Rizos, P. J. Gutierrez, I. M. Navajas and L. M. Lara acknowledge financial support from project PID2021-126365NB-C21 (MCI/AEI/FEDER, UE) and from the Severo Ochoa grant CEX2021-001131-S funded by MCIN/AEI/10.13039/501100011033. J. L. Rizos acknowledges support from the Ministry of Science and Innovation under the funding of the European Union NextGeneration EU/PRTR. J. L. Ortiz acknowledges the Spanish projects PID2020-112789GB-I00 (AEI) and Proyecto de Excelencia de la Junta de Andalucía PY20-01309.
\end{acknowledgements}
\bibliography{main}

\begin{thebibliography}{33}
\expandafter\ifx\csname natexlab\endcsname\relax\def\natexlab#1{#1}\fi

\bibitem[{{Aznar Macias} {et~al.}(2018){Aznar Macias}, {Predatu}, {Vaduvescu},
  \& {Oey}}]{Aznar2018}
{Aznar Macias}, A., {Predatu}, M., {Vaduvescu}, O., \& {Oey}, J. 2018, arXiv
  e-prints, arXiv:1801.09420

\bibitem[{{Barnouin} {et~al.}(2019){Barnouin}, {Daly}, {Palmer}, {Gaskell},
  {Weirich}, {Johnson}, {Al Asad}, {Roberts}, {Perry}, {Susorney}, {Daly},
  {Bierhaus}, {Seabrook}, {Espiritu}, {Nair}, {Nguyen}, {Neumann}, {Ernst},
  {Boynton}, {Nolan}, {Adam}, {Moreau}, {Rizk}, {Drouet D'Aubigny}, {Jawin},
  {Walsh}, {Michel}, {Schwartz}, {Ballouz}, {Mazarico}, {Scheeres}, {McMahon},
  {Bottke}, {Sugita}, {Hirata}, {Hirata}, {Watanabe}, {Burke}, {Dellagiustina},
  {Bennett}, {Lauretta}, {The Osiris-Rex Team}, {Highsmith}, {Small},
  {Vokrouhlick{\'y}}, {Bowles}, {Brown}, {Donaldson Hanna}, {Warren}, {Brunet},
  {Chicoine}, {Desjardins}, {Gaudreau}, {Haltigin}, {Millington-Veloza},
  {Rubi}, {Aponte}, {Gorius}, {Lunsford}, {Allen}, {Grindlay}, {Guevel},
  {Hoak}, {Hong}, {Schrader}, {Bayron}, {Golubov}, {S{\'a}nchez}, {Stromberg},
  {Hirabayashi}, {Hartzell}, {Oliver}, {Rascon}, {Harch}, {Joseph}, {Squyres},
  {Richardson}, {Emery}, {McGraw}, {Ghent}, {Binzel}, {Asad}, {Johnson},
  {Philpott}, {Susorney}, {Cloutis}, {Hanna}, {Connolly}, {Ciceri},
  {Hildebrand}, {Ibrahim}, {Breitenfeld}, {Glotch}, {Rogers}, {Clark},
  {Ferrone}, {Thomas}, {Campins}, {Fernandez}, {Chang}, {Cheuvront}, {Trang},
  {Tachibana}, {Yurimoto}, {Brucato}, {Poggiali}, {Pajola}, {Dotto}, {Epifani},
  {Crombie}, {Lantz}, {Izawa}, {de Leon}, {Licandro}, {Garcia}, {Clemett},
  {Thomas-Keprta}, {van Wal}, {Yoshikawa}, {Bellerose}, {Bhaskaran}, {Boyles},
  {Chesley}, {Elder}, {Farnocchia}, {Harbison}, {Kennedy}, {Knight},
  {Martinez-Vlasoff}, {Mastrodemos}, {McElrath}, {Owen}, {Park}, {Rush},
  {Swanson}, {Takahashi}, {Velez}, {Yetter}, {Thayer}, {Adam}, {Antreasian},
  {Bauman}, {Bryan}, {Carcich}, {Corvin}, {Geeraert}, {Hoffman}, {Leonard},
  {Lessac-Chenen}, {Levine}, {McAdams}, {McCarthy}, {Nelson}, {Page},
  {Pelgrift}, {Sahr}, {Stakkestad}, {Stanbridge}, {Wibben}, {Williams},
  {Williams}, {Wolff}, {Hayne}, {Kubitschek}, {Barucci}, {Deshapriya},
  {Fornasier}, {Fulchignoni}, {Hasselmann}, {Merlin}, {Praet}, {Bierhaus},
  {Billett}, {Boggs}, {Buck}, {Carlson-Kelly}, {Cerna}, {Chaffin}, {Church},
  {Coltrin}, {Daly}, {Deguzman}, {Dubisher}, {Eckart}, {Ellis}, {Falkenstern},
  {Fisher}, {Fisher}, {Fleming}, {Fortney}, {Francis}, {Freund}, {Gonzales},
  {Haas}, {Hasten}, {Hauf}, {Hilbert}, {Howell}, {Jaen}, \&
  {Jayakody}}]{Barnouin2019}
{Barnouin}, O.~S., {Daly}, M.~G., {Palmer}, E.~E., {et~al.} 2019, Nature
  Geoscience, 12, 247

\bibitem[{{Dellagiustina} {et~al.}(2019){Dellagiustina}, {Emery}, {Golish},
  {Rozitis}, {Bennett}, {Burke}, {Ballouz}, {Becker}, {Christensen}, {Drouet
  D'Aubigny}, {Hamilton}, {Reuter}, {Rizk}, {Simon}, {Asphaug}, {Bandfield},
  {Barnouin}, {Barucci}, {Bierhaus}, {Binzel}, {Bottke}, {Bowles}, {Campins},
  {Clark}, {Clark}, {Connolly}, {Daly}, {Leon}, {Delbo'}, {Deshapriya},
  {Elder}, {Fornasier}, {Hergenrother}, {Howell}, {Jawin}, {Kaplan}, {Kareta},
  {Le Corre}, {Li}, {Licandro}, {Lim}, {Michel}, {Molaro}, {Nolan}, {Pajola},
  {Popescu}, {Garcia}, {Ryan}, {Schwartz}, {Shultz}, {Siegler}, {Smith},
  {Tatsumi}, {Thomas}, {Walsh}, {Wolner}, {Zou}, {Lauretta}, \& {Osiris-Rex
  Team}}]{Dellagiustina2019}
{Dellagiustina}, D.~N., {Emery}, J.~P., {Golish}, D.~R., {et~al.} 2019, Nature
  Astronomy, 3, 341

\bibitem[{{Durech} {et~al.}(2010){Durech}, {Sidorin}, \&
  {Kaasalainen}}]{Durech2010}
{Durech}, J., {Sidorin}, V., \& {Kaasalainen}, M. 2010, \aap, 513, A46

\bibitem[{{Fairbairn}(2005)}]{Fairbairn2005}
{Fairbairn}, M.~B. 2005, \jrasc, 99, 92

\bibitem[{Gaskell(2021)}]{Gaskell2021Eros}
Gaskell, R.~W. 2021, {Gaskell Eros Shape Model V1.1},
  \url{https://doi.org/10.26033/d0gq-9427}

\bibitem[{{Golish} {et~al.}(2021){Golish}, {DellaGiustina}, {Li}, {Clark},
  {Zou}, {Smith}, {Rizos}, {Hasselmann}, {Bennett}, {Fornasier}, {Ballouz},
  {Drouet d'Aubigny}, {Rizk}, {Daly}, {Barnouin}, {Philpott}, {Al Asad},
  {Seabrook}, {Johnson}, \& {Lauretta}}]{Golish2021a}
{Golish}, D.~R., {DellaGiustina}, D.~N., {Li}, J.~Y., {et~al.} 2021, \icarus,
  357, 113724

\bibitem[{Hapke(2012)}]{Hapke2012}
Hapke, B. 2012, Theory of Reflectance and Emittance Spectroscopy, 2nd edn.
  (Cambridge University Press)

\bibitem[{{Hergenrother} {et~al.}(2013){Hergenrother}, {Nolan}, {Binzel},
  {Cloutis}, {Barucci}, {Michel}, {Scheeres}, {d'Aubigny}, {Lazzaro},
  {Pinilla-Alonso}, {Campins}, {Licandro}, {Clark}, {Rizk}, {Beshore}, \&
  {Lauretta}}]{Hergenrother2013}
{Hergenrother}, C.~W., {Nolan}, M.~C., {Binzel}, R.~P., {et~al.} 2013, \icarus,
  226, 663

\bibitem[{{Jehl} {et~al.}(2008){Jehl}, {Pinet}, {Baratoux}, {Daydou},
  {Chevrel}, {Heuripeau}, {Manaud}, {Cord}, {Rosemberg}, {Neukum}, {Gwinner},
  {Scholten}, {Hoffman}, {Roatsch}, \& {HRSC Team}}]{Jehl2008}
{Jehl}, A., {Pinet}, P., {Baratoux}, D., {et~al.} 2008, \icarus, 197, 403

\bibitem[{{Kaasalainen} \& {Torppa}(2001)}]{Kaasalainen2001a}
{Kaasalainen}, M. \& {Torppa}, J. 2001, \icarus, 153, 24

\bibitem[{{Kaasalainen} {et~al.}(2001){Kaasalainen}, {Torppa}, \&
  {Muinonen}}]{Kaasalainen2001b}
{Kaasalainen}, M., {Torppa}, J., \& {Muinonen}, K. 2001, \icarus, 153, 37

\bibitem[{{Keane} {et~al.}(2022){Keane}, {Porter}, {Beyer}, {Umurhan},
  {McKinnon}, {Moore}, {Spencer}, {Stern}, {Bierson}, {Binzel}, {Hamilton},
  {Lisse}, {Mao}, {Protopapa}, {Schenk}, {Showalter}, {Stansberry}, {White},
  {Verbiscer}, {Parker}, {Olkin}, {Weaver}, \& {Singer}}]{Keane2022}
{Keane}, J.~T., {Porter}, S.~B., {Beyer}, R.~A., {et~al.} 2022, Journal of
  Geophysical Research (Planets), 127, e07068

\bibitem[{Lambert(1760)}]{Lambert1760}
Lambert, J.~H. 1760, Photometria sive de mensura et gradibus luminis, colorum
  et umbrae (Augsburg: Eberhard Klett)

\bibitem[{Lauretta(2015)}]{Lauretta2015}
Lauretta, D.~S. 2015, OSIRIS-REx Asteroid Sample-Return Mission (Cham: Springer
  International Publishing), 543--567

\bibitem[{{Li}(2005)}]{Li2005PhD}
{Li}, J.-Y. 2005, PhD thesis, University of Maryland, College Park

\bibitem[{{Li} {et~al.}(2013){Li}, {Le Corre}, {Schr{\"o}der}, {Reddy},
  {Denevi}, {Buratti}, {Mottola}, {Hoffmann}, {Gutierrez-Marques}, {Nathues},
  {Russell}, \& {Raymond}}]{Li2013}
{Li}, J.-Y., {Le Corre}, L., {Schr{\"o}der}, S.~E., {et~al.} 2013, \icarus,
  226, 1252

\bibitem[{{Masoumzadeh} {et~al.}(2015){Masoumzadeh}, {Boehnhardt}, {Li}, \&
  {Vincent}}]{Masoumzadeh2015}
{Masoumzadeh}, N., {Boehnhardt}, H., {Li}, J.-Y., \& {Vincent}, J.~B. 2015,
  \icarus, 257, 239

\bibitem[{{Minnaert}(1941)}]{Minnaert1941}
{Minnaert}, M. 1941, \apj, 93, 403

\bibitem[{Möller \& Trumbore(1997)}]{moller1997}
Möller, T. \& Trumbore, B. 1997, Journal of Graphics Tools, 2, 21

\bibitem[{{Ortiz} {et~al.}(2017){Ortiz}, {Santos-Sanz}, {Sicardy},
  {Benedetti-Rossi}, {B{\'e}rard}, {Morales}, {Duffard}, {Braga-Ribas}, {Hopp},
  {Ries}, {Nascimbeni}, {Marzari}, {Granata}, {P{\'a}l}, {Kiss}, {Pribulla},
  {Kom{\v{z}}{\'\i}k}, {Hornoch}, {Pravec}, {Bacci}, {Maestripieri}, {Nerli},
  {Mazzei}, {Bachini}, {Martinelli}, {Succi}, {Ciabattari}, {Mikuz},
  {Carbognani}, {Gaehrken}, {Mottola}, {Hellmich}, {Rommel},
  {Fern{\'a}ndez-Valenzuela}, {Campo Bagatin}, {Cikota}, {Cikota}, {Lecacheux},
  {Vieira-Martins}, {Camargo}, {Assafin}, {Colas}, {Behrend}, {Desmars},
  {Meza}, {Alvarez-Candal}, {Beisker}, {Gomes-Junior}, {Morgado}, {Roques},
  {Vachier}, {Berthier}, {Mueller}, {Madiedo}, {Unsalan}, {Sonbas}, {Karaman},
  {Erece}, {Koseoglu}, {Ozisik}, {Kalkan}, {Guney}, {Niaei}, {Satir},
  {Yesilyaprak}, {Puskullu}, {Kabas}, {Demircan}, {Alikakos}, {Charmandaris},
  {Leto}, {Ohlert}, {Christille}, {Szak{\'a}ts}, {Tak{\'a}csn{\'e} Farkas},
  {Varga-Vereb{\'e}lyi}, {Marton}, {Marciniak}, {Bartczak}, {Santana-Ros},
  {Butkiewicz-B{\k{a}}k}, {Dudzi{\'n}ski}, {Al{\'\i}-Lagoa}, {Gazeas},
  {Tzouganatos}, {Paschalis}, {Tsamis}, {S{\'a}nchez-Lavega},
  {P{\'e}rez-Hoyos}, {Hueso}, {Guirado}, {Peris}, \&
  {Iglesias-Marzoa}}]{Ortiz2017}
{Ortiz}, J.~L., {Santos-Sanz}, P., {Sicardy}, B., {et~al.} 2017, \nat, 550, 219

\bibitem[{Ostro \& Connelly(1984)}]{Ostro1984}
Ostro, S.~J. \& Connelly, R. 1984, Icarus, 57, 443

\bibitem[{{Ostro} {et~al.}(1988){Ostro}, {Connelly}, \& {Dorogi}}]{Ostro1988}
{Ostro}, S.~J., {Connelly}, R., \& {Dorogi}, M. 1988, \icarus, 75, 30

\bibitem[{{Preusker} {et~al.}(2017){Preusker}, {Scholten}, {Matz}, {Roatsch},
  {Hviid}, {Mottola}, {Knollenberg}, {K{\"u}hrt}, {Pajola}, {Oklay}, {Vincent},
  {Davidsson}, {A'Hearn}, {Agarwal}, {Barbieri}, {Barucci}, {Bertaux},
  {Bertini}, {Cremonese}, {Da Deppo}, {Debei}, {De Cecco}, {Fornasier},
  {Fulle}, {Groussin}, {Guti{\'e}rrez}, {G{\"u}ttler}, {Ip}, {Jorda}, {Keller},
  {Koschny}, {Kramm}, {K{\"u}ppers}, {Lamy}, {Lara}, {Lazzarin}, {Lopez
  Moreno}, {Marzari}, {Massironi}, {Naletto}, {Rickman}, {Rodrigo}, {Sierks},
  {Thomas}, \& {Tubiana}}]{Preusker2017}
{Preusker}, F., {Scholten}, F., {Matz}, K.~D., {et~al.} 2017, \aap, 607, L1

\bibitem[{{Rizk} {et~al.}(2018){Rizk}, {Drouet d'Aubigny}, {Golish}, {Fellows},
  {Merrill}, {Smith}, {Walker}, {Hendershot}, {Hancock}, {Bailey},
  {DellaGiustina}, {Lauretta}, {Tanner}, {Williams}, {Harshman}, {Fitzgibbon},
  {Verts}, {Chen}, {Connors}, {Hamara}, {Dowd}, {Lowman}, {Dubin}, {Burt},
  {Whiteley}, {Watson}, {McMahon}, {Ward}, {Booher}, {Read}, {Williams},
  {Hunten}, {Little}, {Saltzman}, {Alfred}, {O'Dougherty}, {Walthall},
  {Kenagy}, {Peterson}, {Crowther}, {Perry}, {See}, {Selznick}, {Sauve},
  {Beiser}, {Black}, {Pfisterer}, {Lancaster}, {Oliver}, {Oquest}, {Crowley},
  {Morgan}, {Castle}, {Dominguez}, \& {Sullivan}}]{Rizk2018}
{Rizk}, B., {Drouet d'Aubigny}, C., {Golish}, D., {et~al.} 2018, \ssr, 214, 26

\bibitem[{{Rizos} {et~al.}(2021){Rizos}, {de Le{\'o}n}, {Licandro}, {Golish},
  {Campins}, {Tatsumi}, {Popescu}, {DellaGiustina}, {Pajola}, {Li}, {Becker},
  \& {Lauretta}}]{Rizos2021}
{Rizos}, J.~L., {de Le{\'o}n}, J., {Licandro}, J., {et~al.} 2021, \icarus, 364,
  114467

\bibitem[{{Rizos} {et~al.}(2024){Rizos}, {Fern{\'a}ndez-Valenzuela}, {Ortiz},
  {Rommel}, {Sicardy}, {Morales}, {Santos-Sanz}, {Leiva}, {Vara-Lubiano},
  {Morales}, {Kretlow}, {Alvarez-Candal}, {Holler}, {Duffard},
  {G{\'o}mez-Lim{\'o}n}, {Desmars}, {Souami}, {Assafin}, {Benedetti-Rossi},
  {Braga-Ribas}, {Camargo}, {Colas}, {Lecacheux}, {Gomes-J{\'u}nior},
  {Vieira-Martins}, {Pereira}, {Morgado}, {Kilic}, {Redfield}, {Soloff},
  {McGregor}, {Green}, {Midavaine}, {Schreurs}, {Lecossois}, {Boninsegna},
  {Ida}, {Le Cam}, {Isobe}, {Watanabe}, {Yuasa}, {Watanabe}, \&
  {Kidd}}]{Rizos2024}
{Rizos}, J.~L., {Fern{\'a}ndez-Valenzuela}, E., {Ortiz}, J.~L., {et~al.} 2024,
  \aap, 689, A82

\bibitem[{{Shepard} {et~al.}(2018){Shepard}, {Timerson}, {Scheeres}, {Benner},
  {Giorgini}, {Howell}, {Magri}, {Nolan}, {Springmann}, {Taylor}, \&
  {Virkki}}]{Shepard2018}
{Shepard}, M.~K., {Timerson}, B., {Scheeres}, D.~J., {et~al.} 2018, \icarus,
  311, 197

\bibitem[{{Showalter} {et~al.}(2021){Showalter}, {Benecchi}, {Buie}, {Grundy},
  {Keane}, {Lisse}, {Olkin}, {Porter}, {Robbins}, {Singer}, {Verbiscer},
  {Weaver}, {Zangari}, {Hamilton}, {Kaufmann}, {Lauer}, {Mehoke}, {Mehoke},
  {Spencer}, {Throop}, {Parker}, {Stern}, {New Horizons Geology}, \&
  Team}]{Showalter2021}
{Showalter}, M.~R., {Benecchi}, S.~D., {Buie}, M.~W., {et~al.} 2021, \icarus,
  356, 114098

\bibitem[{{Takir} {et~al.}(2015){Takir}, {Clark}, {Drouet d'Aubigny},
  {Hergenrother}, {Li}, {Lauretta}, \& {Binzel}}]{Takir2015}
{Takir}, D., {Clark}, B.~E., {Drouet d'Aubigny}, C., {et~al.} 2015, \icarus,
  252, 393

\bibitem[{{Tatsumi} {et~al.}(2021){Tatsumi}, {Popescu}, {Campins}, {de
  Le{\'o}n}, {Garc{\'\i}a}, {Licandro}, {Simon}, {Kaplan}, {DellaGiustina},
  {Golish}, \& {Lauretta}}]{Tatsumi2021}
{Tatsumi}, E., {Popescu}, M., {Campins}, H., {et~al.} 2021, \mnras, 508, 2053

\bibitem[{{Tedesco} {et~al.}(1982){Tedesco}, {Tholen}, \&
  {Zellner}}]{Tedesco1982}
{Tedesco}, E.~F., {Tholen}, D.~J., \& {Zellner}, B. 1982, \aj, 87, 1585

\bibitem[{{Yeomans} {et~al.}(2000){Yeomans}, {Antreasian}, {Barriot},
  {Chesley}, {Dunham}, {Farquhar}, {Giorgini}, {Helfrich}, {Konopliv},
  {McAdams}, {Miller}, {Owen}, {Scheeres}, {Thomas}, {Veverka}, \&
  {Williams}}]{Yeomans2000}
{Yeomans}, D.~K., {Antreasian}, P.~G., {Barriot}, J.~P., {et~al.} 2000,
  Science, 289, 2085

\end{thebibliography}

\begin{appendix}
\onecolumn

\section{Hapke photometric model equations}\label{appendix}

The Hapke model \citep{Hapke2012} is based on radiative transfer theory and provides a more physically grounded framework for describing surface scattering. It accounts for single and multiple scattering, surface roughness, and opposition effects. The $RADF$ in our implementation is computed as:

\begin{equation}
RADF = LS(i, e)  K  \frac{w}{4} \left[ p(\alpha)(1 + B_{S0} B_{S}(\alpha)) + M(i, e) \right] \left[ 1 + B_{C0} B_{C}(\alpha) \right] S(i, e, \psi)
\end{equation}

Here:

\begin{itemize}
\item $LS(i, e)$ is the Lommel-Seeliger function, $\frac{\cos i}{\cos i + \cos e}$
\\

\item $K$ is the porosity factor, $\frac{-\ln (1 - 1.209\phi^{2/3})}{1.209\phi^{2/3}}$, with $\phi$ the filling factor (1 - porosity)
\\

\item $p(\alpha)$ is the phase function, for which we use the Henyey-Greenstein form:
\begin{equation}
\begin{split}
p(\alpha) &= \frac{(1 + c_{HG})}{2} \frac{(1 - b_{HG}^2)}{(1 - 2b_{HG} \cos \alpha + b_{HG}^2)^{3/2}} \
&+ \frac{(1 - c_{HG})}{2}  \frac{(1 - b_{HG}^2)}{(1 + 2b_{HG} \cos \alpha + b_{HG}^2)^{3/2}}
\end{split}
\end{equation}
\\

\item $B_S(\alpha)$ is the Shadow Hiding Opposition Effect (SHOE), $ \frac{1}{1 + \tan(\alpha/2)/h_S}$
\\

\item $M(i, e)$ is the Isotropic Multiple-Scattering Approximation (IMSA):
\begin{equation}
M(i, e) = H\left(\frac{\cos(i)}{K}, w\right) H\left(\frac{\cos(e)}{K}, w\right) - 1
\end{equation}
where $H(x, w)$ is the Ambartsumian-Chandrasekhar function:
\begin{equation}
H(x, w) \approx \left[ 1 - w x \left( r_0 + \frac{1 - 2 r_0 x}{2} \ln \left(\frac{1 + x}{x}\right) \right) \right]^{-1}
\end{equation}
and $r_0$ is the diffusive reflectance:
\begin{equation}
r_0 = \frac{1 - \sqrt{1 - w}}{1 + \sqrt{1 - w}}
\end{equation}
\\

\item $B_C(\alpha)$ is the Coherent Backscatter Opposition Effect (CBOE):
\begin{equation}
B_C(\alpha) = \frac{1 + \frac{1 - \exp(-\tan(\alpha/2) / h_C)}{\tan(\alpha/2) / h_C}}{2 (1 + \tan(\alpha/2) / h_C)^2}
\end{equation}
\\

\item The shadowing function \( S(i, e, \psi) \) accounts for the effect of macroscopic surface roughness. Its formulation varies depending on the relative values of the incidence angle \( i \) and the emission angle \( e \):

\paragraph{Case 1: \( i \leq e \)}

\begin{equation}
S(i, e, \psi) = \frac{\mu}{\eta(e)}  \frac{\mu_0}{\eta(i)}  \frac{\chi(\theta_p)}{1 - f(\psi) + f(\psi) \chi(\theta_p) \left( \frac{\mu_0}{\eta(i)} \right)}
\end{equation}

with

\begin{align}
\mu_0 &= \chi(\theta_p) \left[ \cos(i) + \sin(i) \tan(\theta_p) \frac{\cos(\psi) E_2(e) + \sin^2(\psi/2) E_2(i)}{2 - E_1(e) - (\psi/\pi) E_1(i)} \right] \\
\mu &= \chi(\theta_p) \left[ \cos(e) + \sin(e) \tan(\theta_p)  \frac{E_2(e) - \sin^2(\psi/2) E_2(i)}{2 - E_1(i) - (\psi/\pi) E_1(i)} \right]
\end{align}

\paragraph{Case 2: \( e < i \)}

\begin{equation}
S(i, e, \psi) = \frac{\mu}{\eta(e)}  \frac{\mu_0}{\eta(i)}  \frac{\chi(\theta_p)}{1 - f(\psi) + f(\psi) \chi(\theta_p) \left( \frac{\mu}{\eta(e)} \right)}
\end{equation}

with

\begin{align}
\mu_0 &= \chi(\theta_p) \left[ \cos(i) + \sin(i) \tan(\theta_p)  \frac{E_2(i) - \sin^2(\psi/2) E_2(e)}{2 - E_1(i) - (\psi/\pi) E_1(e)} \right] \\
\mu &= \chi(\theta_p) \left[ \cos(e) + \sin(e) \tan(\theta_p)  \frac{\cos(\psi) E_2(i) + \sin^2(\psi/2) E_2(e)}{2 - E_1(i) - (\psi/\pi) E_1(e)} \right]
\end{align}

The terms involved in the expressions above are defined as follows:

\begin{align}
\chi(\theta_p) &= \frac{1}{\sqrt{1 + \pi \tan^2(\theta_p)}} \\
\eta(y) &= \chi(\theta_p) \left[ \cos(y) + \sin(y) \tan(\theta_p) \frac{E_2(y)}{2 - E_1(y)} \right] \\
E_1(y) &= \exp \left[ -\frac{2}{\pi} \cot(\theta_p) \cot(y) \right] \\
E_2(y) &= \exp \left[ -\frac{1}{\pi} \cot^2(\theta_p) \cot^2(y) \right] \\
f(\psi) &= \exp \left[ -2 \tan \left( \frac{\psi}{2} \right) \right]
\end{align}

\end{itemize}

\section{OSIRIS-REx Bennu's image}

\begin{figure}[ht]
    \centering
    \includegraphics[width=0.45\textwidth]{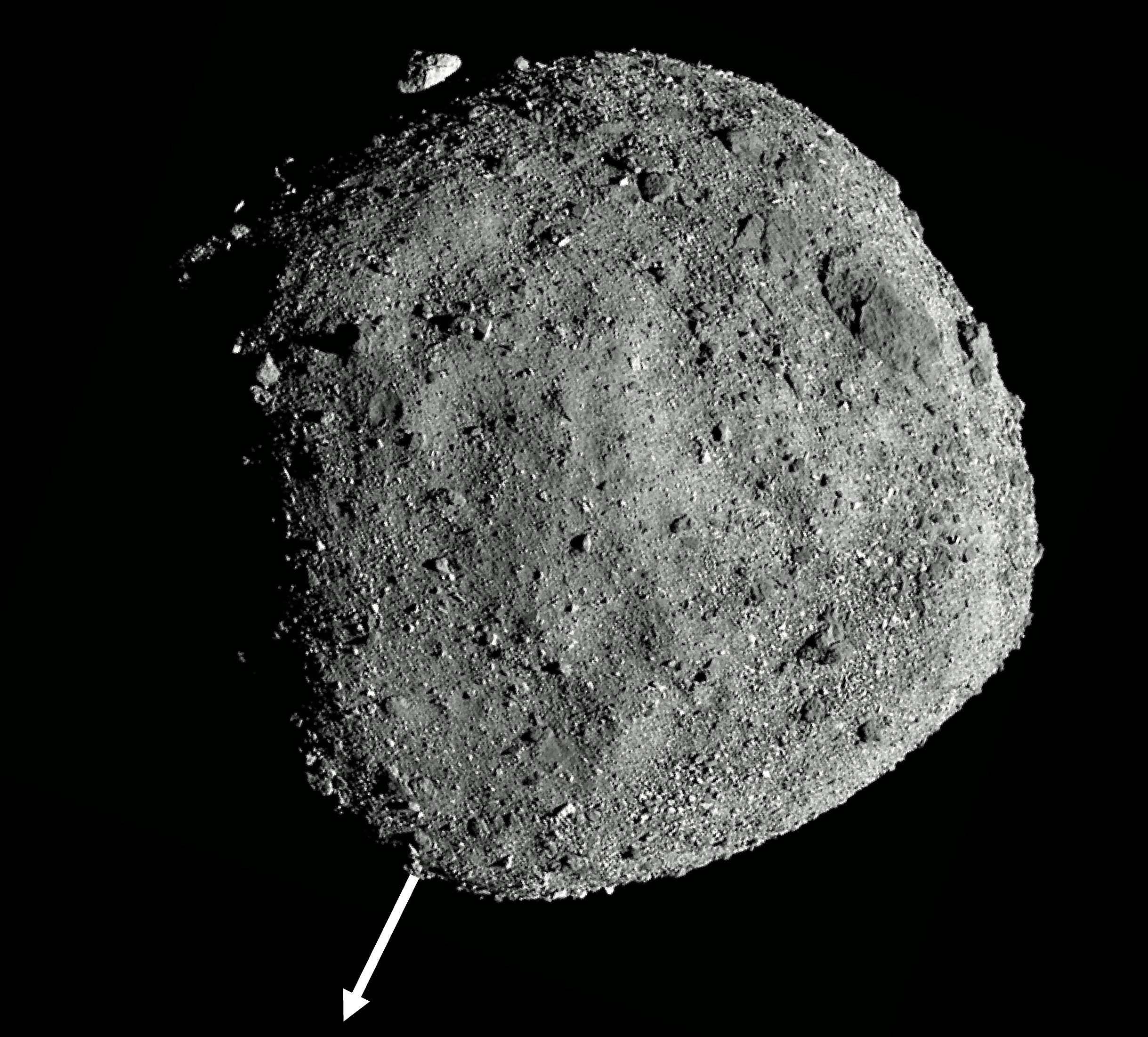}
    \caption{Bennu imaged by OSIRIS-REx (PolyCam) during the approach phase. The image has been rotated so that the spin axis orientation approximates the aspect as seen from Earth in September 2005. Credit: NASA/Goddard/University of Arizona.}
    \label{fig:bennu_polycam}
\end{figure}

To provide visual context, we include in Figure~\ref{fig:bennu_polycam} a global view of Bennu’s surface acquired by the OSIRIS-REx spacecraft using the PolyCam imager during its approach. This mosaic offers a comprehensive look at Bennu’s topography and albedo variations across its entire surface. The image was rotated to approximately match the viewing geometry from Earth on 14--17 September 2005, the period for which we generated synthetic light curves for validation against ground-based data. This orientation helps illustrate the correspondence between surface features and their potential photometric signature in rotational brightness variations.

Although the orientation does not exactly reproduce the sub-Earth latitude or illumination conditions of the 2005 observations, it serves as a valuable qualitative reference for assessing discrepancies between the synthetic simulations and the actual observational data.

\end{appendix}

\end{document}